\documentclass[runningheads]{llncs}
\usepackage[T1]{fontenc}
\usepackage{booktabs}
\usepackage{graphicx}
\usepackage{makecell}
\usepackage{xcolor}
\usepackage{float}
\usepackage{linguex}

\begin{document}
%
%\title{Learning to Intervene in Classroom Hate Incidents with AI Assistant }
\title{Leveraging a Multi-Agent LLM-Based  System to Educate Teachers in Hate Incidents Management}
% Beyond Workshops: Training Management of Hate Incidents with Agentic LLMs
% Rethinking Teacher Training: Leveraging Agentic AI to Empower Teachers in Handling Hate Incidents
% Empowering Educators to Manage Hate Incidents Using LLMs-Based Multi-Agent Systems

\titlerunning{A Multi-Agent LLM-Based System for Hate Incidents Management}
%
%\author{The Author(s)}
\author{Ewelina Gajewska\inst{1}\orcidID{0009-0006-6012-4787} \and
Michal Wawer\orcidID{0009-0004-2717-1616} \and
Katarzyna Budzynska\inst{1}\orcidID{0000-0001-9674-9902} \and
Jaroslaw A. Chudziak\inst{1}\orcidID{0000-0003-4534-8652}
}
\authorrunning{E. Gajewska et al.}
%\authorrunning{The Author(s)}


\institute{Warsaw University of Technology, 00-661 Warsaw, Poland \\
\email{ewelina.gajewska.dokt@pw.edu.pl}
}

\maketitle              % typeset the header of the contribution

\begin{abstract}
Computer-aided teacher training is a state-of-the-art method designed to enhance teachers' professional skills effectively while minimising concerns related to costs, time constraints, and geographical limitations. We investigate the potential of large language models (LLMs) in teacher education, using a case of teaching hate incidents management in schools. 
To this end, we create a multi-agent LLM-based system that mimics realistic situations of hate, using a combination of retrieval-augmented prompting and persona modelling. It is designed to identify and analyse hate speech patterns, predict potential escalation, and propose effective intervention strategies. By integrating persona modelling with agentic LLMs, we create contextually diverse simulations of hate incidents, mimicking real-life situations. 
The system allows teachers to analyse and understand the dynamics of hate incidents in a safe and controlled environment, providing valuable insights and practical knowledge to manage such situations confidently in real life. 
Our pilot evaluation demonstrates teachers' enhanced understanding of the nature of annotator disagreements and the role of context in hate speech interpretation, leading to the development of more informed and effective strategies for addressing hate in classroom settings. 
% krócej 

\end{abstract}

% dopisac ewaluacje
% contextually diverse - w ewaluacji uwzlednic, focus na to
% requirements od nauczycieli (pod kolejną wersje systemu) i potem ponowna ewaluacja z nauczyielami-uzytkownikami
% management skills spięte z challenges
% dodatkowe pytania do ewaluacji?
% \newpage
% Moreover, dealing with hate incidents in real-life can be emotionally draining for teachers who may feel overwhelmed or stressed by the harmful behaviour they witness.  - do wstepu
% Leveraging a Multi-Agent LLM-Based System to Educate Teachers in  Hate Incident Management

%\section{Introduction}
% 90
% scenariusz w intro hate sytuacji jaka moze wystapic - o jaki rodzaj hate nam chodzi / takie summary artykulu (jakie na prezentacji by mogly byc) 
% spis obrazow w artykule 

%\section{Related Works}
% 90
% ladny obrazek z jakiejs konf/artykulu o naszym podejsciu
% \subsection{AI in Education}
% \subsection{LLMs as Intelligent Cognitive Systems}
% \subsection{The Potential of LLMs in Education}
% \section{Managing Hate Incidents}
% 90
% \subsection{Challenges}
% \subsection{Solution}% oczekiwania codo rozwiazania
% \subsection{Evaluation strategy} % ilosciowe accuracy i jakosciowe a badan uzytecznosci uzytkowników
% \section{XYZ: Platform for Hate Incident Management}
% 30
% \subsection{Architecture}
% \subsection{Data}
% \subsection{Evaluation}
% \section{Application of XYZ in Teacher Education }% case based teacher education - / 5.1 & 5.2
%\section{Results}
% \subsection{Annotation: }
% \subsection{Workshop}
% 0
% \section{Discussion}
% 0
% \section{Conclusions}
% 0

\section{Introduction}
Early researchers and educators recognised the potential of artificial intelligence (AI) to revolutionise learning and training. In recent years,  virtual classrooms, AI-powered tutoring systems, and interactive learning platforms have become commonplace. 
%These technologies have been embraced to enhance student learning, offering personalised and accessible educational experiences.
Despite these advancements, the widespread adoption of AI in educational training remains limited. 
Recent advancements in language technologies have introduced new opportunities to create accessible and highly interactive educational experiences.
We propose to leverage these technologies to create new educational opportunities for teachers to learn hate incident management skills. The high prevalence of hate in recent years has highlighted the urgent need for designing effective intervention strategies \cite{stefuanitua2021hate}, and teachers, who play a pivotal role in shaping the social and emotional environment of their classrooms, are often on the front lines when such incidents occur. 

Consequently, it is important to provide teachers with comprehensive training on how to identify, address, and prevent hate incidents to create a safe and inclusive environment for students. This training should not only equip teachers with the necessary skills to manage these situations effectively but also foster a deeper understanding of the underlying issues that contribute to hate-based behaviour. 
Trained on massive amounts of data, LLMs harness the power of deep learning to generate human-like responses, provide accurate information, and help individuals complete complex tasks across a variety of domains \cite{achiam2023gpt}. 
A recent review of the transformative potential of AI emphasises LLMs's ability to enhance task proficiency, foster creativity in writing, and augment student engagement during learning sessions \cite{alshahrani2024review,raiaanReviewOfLLMs2024}. 
LLMs have the potential to revolutionise also medical education by acting as interactive virtual tutors \cite{mondal2023chatgpt}. 

A number of articles recognise the potential of using AI tools like ChatGPT in teacher training programs to address the disconnection between teachers' theoretical knowledge and real-world teaching applications, often in the form of case-based reasoning \cite{pitura2024vr,kim2024leading}. 
To this end, we explore the potential of LLM-based systems as practice-oriented spaces in which teachers can learn to effectively manage hate incidents. 
Specifically, we propose a multi-agent LLM-based system equipped with a custom personalisation module that is designed to identify hate incidents while being sensitive to cultural differences. 
%We critically examine key advantages of such AI systems for teacher education over traditional training methods and ways these systems can complement existing training programs. 
%By acknowledging the limitations of these systems, such as potential biases in the models, we ensure a balanced perspective on the topic. 
We aim to demonstrate how such systems can complement traditional training programs, emphasising the importance of integrating human expertise and ethical considerations. Through this comprehensive analysis, we seek to offer valuable insights into the integration of innovative technologies in educational settings.

% bez podrozdzialów
\section{Managing Hate Incidents}

%\subsection{Challenges}
Psychological studies link experience of hateful attacks to both short and long-term effects such as mood swings, emotional harm and erosion of social trust \cite{stefuanitua2021hate}. 
However, identification of hateful speech can be highly subjective by evoking different reactions and interpretations depending on the individual identity \cite{ross2016measuring,sap2021annotators}. 
%These differences come in part from the context-dependent nature of hate and the subtlety of language constructions such as irony and sarcasm often employed in hateful attacks. 
These differences are reflected in the lack of uniform standards of management of hate speech cases as well as differing definitions and policies on hate incidents. 
On the other hand, from 25\% to over 50\% of students witness at least one hate incident in school over the period of one year and over 10\% of students reported being a victim of hate \cite{kansok2023systematic}. Furthermore, over 60\% of teachers witnessed a situation where students made publicly offensive comments about the skin colour, sexual orientation, religion, or gender of a group of people. 
Yet, only 40\% of the teachers expressed the intention to intervene in such situations. 
Moreover, less professionally experienced teachers have been found to employ different intervention strategies than more experienced teachers \cite{bilz2024teachers}.
%Teachers must consider the complexities of students’ diverse backgrounds and sensitivities, which can make it challenging to respond to hate incidents in a way that is effective and respectful for all students.
Schools provide few educational opportunities for teachers on how to handle hate incidents, leaving them unprepared to address these situations effectively.

%\subsection{Solution}
%To this end, we utilise the potential of agentic LLMs as a valuable tool in teacher education, using as a use case of LLM-based workshop  handle hate incidents by scenario-based training. 
Despite the advancements in educational technology, the use of AI in teacher training remains relatively underexplored. Traditional methods such as seminars and in-person training with human experts still dominate the field. These methods, while valuable, have limitations in terms of scalability, personalisation, and continuous learning opportunities. 
%Workshops typically require significant resources, including time, money, and personnel, and they can only accommodate a limited number of participants at a time. 
%In contrast, LLMs can scale to reach a vast number of teachers simultaneously, regardless of their geographical location. They allow teachers to continually refine their responses to hate incidents, offering personalised training experiences adapted to individual needs. Due to the sensitive nature of the topic, LLM-based training seems to be also a better option in providing a safe and controlled environment for training as dealing with real hate incidents can be emotionally taxing for teachers, who may feel overwhelmed or stressed by the harmful behaviour they witness. 
To this end, we design a multi-agent LLM-based platform with the purpose of supporting teachers in managing hate speech in school settings. The system is particularly valuable for understanding the dynamics of hate incidents in order to program effective intervention strategies, providing continuous, scalable, and personalised training opportunities - the key features of educational AI systems \cite{oyekoya2021exploring}. 
We pose four research questions that regard the development of such a platform: 
\textbf{RQ1.} How does a machine compare to humans in annotating hate speech? 
\textbf{RQ2.} How does the personalisation of agents affect the annotation of hate speech? 
\textbf{RQ3.} How can the proposed platform be utilised in educational workshops, and in which areas does this platform have the greatest potential for utilisation? 
\textbf{RQ4.} What are the criteria for evaluating the usefulness and effectiveness of such a platform in an educational context?

To assess the efficacy of our system, we devise a two-fold evaluation strategy, encompassing both the capability of AI agents to recognise hate speech and the effectiveness of integrating these agents into an interactive system that supports teachers in real time. 
First, state-of-the-art LLM, GPT-o1-mini, is selected as the basis of our platform. Its performance is assessed with two benchmark datasets: HateXplain \cite{mathew2021HateXplain}, which includes explicit hate speech examples, and Latent Hatred \cite{elsherief-etal-2021-latent}, which presents more nuanced, implicit, and context-dependent hate speech cases. 
This evaluation allows us to answer RQ.1 and 2. 
The second part of our evaluation investigates the practical utility of embedding AI agents into an interactive system designed for teachers. This evaluation aims to determine whether such a system enhances teachers' preparedness and decision-making in real-world scenarios. 
The evaluation is conducted with a system where teachers can input descriptions of hate incidents and receive structured responses. The agents analyse each case, simulating diverse student perspectives, and provide explanations regarding the nature of hate speech, potential escalation risks, and suggested intervention strategies. This allows us to answer RQ.3 and 4.

% ARISE: (Agent Resource for Incident Support and Education)
% arise jako ogolna platforma, a tutaj jako jeden przyklad do anotacji hate speechu - miedzy 4 i 4.1, np. rozkodowanie ARISE w odpowiedzi na xyz, proponujemy ..., uzasadnienie dlaczego 
\section{ARISE: Agent Resource for Incident Support and Education} \label{sec:methodology}
This section presents the architecture of the proposed ARISE platform. We provide details on the LLMs employed in the study, characteristics of agents, the hate speech analysis workflow, and evaluation of the classification performance.

\subsection{Architecture}
%The architecture of our platform is designed to facilitate interactive and context-aware support for teachers managing hate incidents in schools. 
To explore the impact of external knowledge on model performance, we employed two configurations. The first utilised Retrieval-Augmented Generation (RAG) \cite{lewis2021retrievalaugmentedgeneration}, providing supplementary materials such as academic articles, case studies, and official definitions of hate speech to enhance contextual understanding. The second configuration relied solely on the intrinsic capabilities of the pre-trained models. % without additional contextual support.
Furthermore, we examined the influence of agent collaboration \cite{harbarChudziakOxfordStyleDebatesLLMs} by employing two experimental setups. In the single-agent setup, three independent AI agents, each simulating a university student from different academic disciplines, performed separate analyses. In contrast, the multi-agent setup introduced a manager agent, simulating an academic professor, to facilitate collaboration among the student agents. %The manager aggregated insights, enabling more refined interpretations. 
The results from these configurations provided insights into the models accuracy in detecting hate speech and the potential benefits of multi-agent collaboration.

\begin{figure}[h]
    \centering
    \includegraphics[width=1\linewidth]{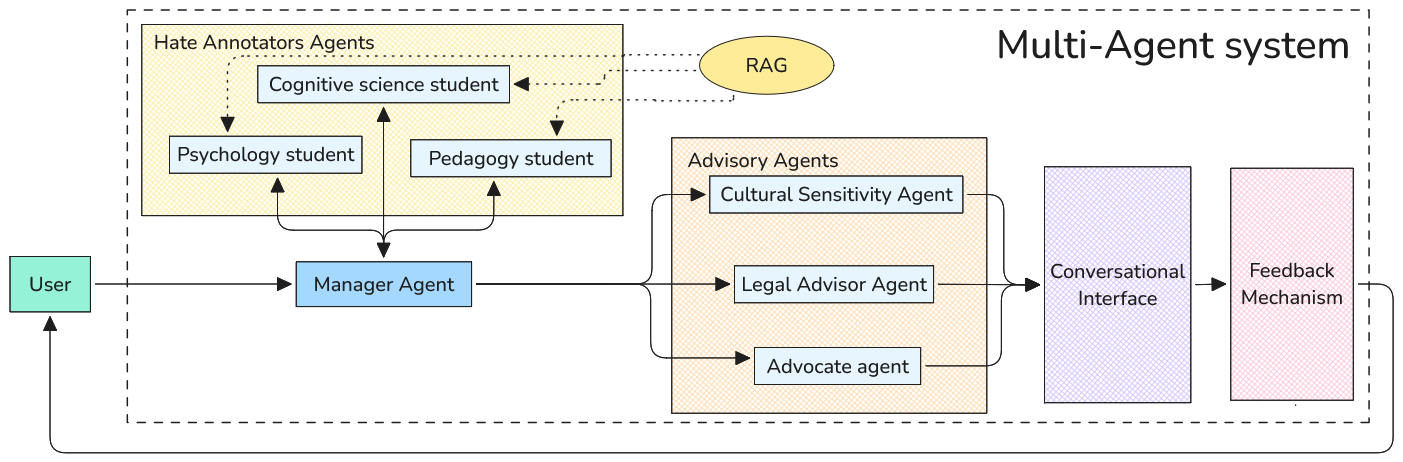}
    \caption{Architecture of the ARISE platform.} \label{fig:architecture}
\end{figure}

The system uses CrewAI framework, which implements student-role and domain-specific advisors AI agents and a conversational graphical interface to ensure dynamic and adaptive interaction with user. 
At the core of the system, a manager agent orchestrates the flow of information and decision-making, ensuring structured analysis. Teachers interact with the system through the manager agent (see Fig. \ref{fig:architecture}), which coordinates responses from different sub-agents specialised in distinct aspects of hate speech analysis and intervention.
The primary components include: \textbf{(1)  Manager Agent}  acts as a supervisor, facilitating collaboration among student agents. Modelled as an professor, it aggregates insights, resolves conflicts, and refines recommendations for intervention strategies. \textbf{(2)  Student Agents}  simulate students with different academic backgrounds, including psychology, pedagogy, and cognitive science. Each student agent analyses incidents from its respective disciplinary perspective, ensuring diverse interpretations of hate speech cases. \textbf{(3)  Advisory Agents} analyses incidents through diverse cultural lenses, examining how different cultural backgrounds and lived experiences might affect the interpretation of potentially hateful content. At the end, these agents provide feedback to the user. \textbf{(4)  RAG Module} – We incorporated dynamic retrieval of relevant contextual materials, such as case studies, legal definitions, and academic research. %, to enhance the AI agents' understanding and decision-making capabilities.
\textbf{(5) Conversational Interface} – Teachers interact with the system through a user-friendly chat interface, where they can describe incidents, ask questions, and receive real-time feedback. %The interface provides a structured yet flexible means of engaging with AI agents for training and decision support.

%The architecture follows a modular approach to facilitate scalability and adaptability. The multi-agent framework ensures robust decision-making by allowing different perspectives to be considered simultaneously. The RAG module further augments the agents capabilities, ensuring that responses are well-grounded in established knowledge. %By integrating these components, the system provides an interactive, context-aware, and adaptive tool for teacher training and support in hate incident management.

% \begin{figure}[h]
%     \centering
%     \includegraphics[width=0.82\linewidth]{agent_process.png}
%     \caption{A procedure of analysing hate speech with our system.} \label{fig:agents_process}
% \end{figure}

%\subsection{Multi-Agent System}
%overview of the idea - whether single or multi agent system, what we mean by that, why multi and not single agent platform or vice versa

%\begin{figure}
 %   \centering
 %   \includegraphics[width=0.97\linewidth]{agent_model.png}
 %   \caption{Caption} \label{fig:masmodel}
%\end{figure}

\subsection{LLM-based Agents}

Our system employs three specialised student agents, each with a unique backstory and perspective that influences their approach to hate speech detection. These personas are designed to leverage different academic disciplines and personal experiences, enabling a more comprehensive analysis of potential hate incidents. 
% Each agent's backstory serves as a foundation for their analytical approach. The Psychology student, motivated by personal experience with online harassment, brings expertise in social psychology and group dynamics to understand the emotional impact and underlying intentions of potentially hateful communications. The Pedagogy student, shaped by early exposure to community education, excels at identifying subtle forms of prejudice embedded in seemingly neutral language, drawing from their understanding of how educational environments can either perpetuate or combat discrimination. The Cognitive science student combines computational expertise with insights into language processing, offering a unique perspective on how hate speech manifests in complex linguistic patterns.
Backstories are not merely narrative elements but are integrated into each agent's prompt engineering \cite{kostkaChudziakLogiccalReasoning2024,zhoSurveyOfLLMsAgents2023}, influencing how they process and analyse potential hate incidents. By maintaining consistent personas across interactions, the agents provide reliable and specialised perspectives that complement each other in the analysis process \cite{cinkusz2024towardsLlmAugmentedMultiagentSystems}. This approach allows for a multi-faceted examination of hate speech incidents, combining emotional intelligence, educational insight, and technical analysis in a single system.

\subsection{Evaluation of Classification Performance}
%\subsubsection{Quantitative Analysis}
% All three LLM students identify hate speech with accuracy exceeding previously proposed BERT, T5, GTP-2 models and Perspective API \cite{mathew2021HateXplain,kim2022hate,he2024you}.

We use 100 samples from each dataset to evaluate the performance of different agent configurations and models. Results are presented in Table \ref{tab:results-gpt}. 
%We conducted testing of our agents using 100 samples from each dataset, to evaluate the performance of different agent configurations and models. 
The accuracy of detecting explicit hate speech cases from the HateXplain dataset ranges from 72\% to 79\%. The implementation of RAG improves performance across most configurations, with an average increase of 4.5 p.p.. The multi-agent setup combined with RAG achieves the highest accuracy (79\%), suggesting that collaborative analysis can enhance detection accuracy and outperform single-agent setups. %Notably, even without RAG, the multi-agent configuration for GPT-o1-mini achieved 78\% accuracy, outperforming single-agent setups.
The Latent Hatred dataset presents a more complex challenge, requiring models to classify implicit forms of hate speech into seven distinct categories: grievance, incitement, stereotypes, inferiority, irony, threats and the other category. In this task, the performance of GPT-o1-mini ranges from 47\% to 60\%, substantially lower than for detecting explicitly hateful speech. Again, the multi-agent setup consistently outperforms single-agent configurations, reaching 60\% accuracy. The RAG implementation provides improvements of 1-7 p.p. in classification accuracy. 
%\vspace{-0.3cm}

\begin{table}[h!]
    \caption{Performance of GPT-o1-mini across different student types and setups.}
    \centering
    \resizebox{\textwidth}{!}{
        \begin{tabular}{|l|r|r|r|r|}
            \hline
            gpt-o1-mini & \makecell{Psychology \\ student} & \makecell{Pedagogy \\ student} & \makecell{Cognitive science \\ student} & \makecell{Mixture of \\ agents} \\ \hline
            HateXplain w/o RAG & 72\% & 76\% & 74\% & 78\% \\
            HateXplain w/ RAG & 79\% & 78\% & 79\% & 79\% \\
            Implicit-Hate w/o RAG & 51\% & 49\% & 47\% & 55\% \\
            Implicit-Hate w/ RAG & 55\% & 56\% & 53\% & 60\% \\ \hline
        \end{tabular}%
    }    
    \label{tab:results-gpt}
\end{table}

%\begin{figure}
 %   \centering
  %  \includegraphics[width=0.97\linewidth]{ui_design.png}
  %  \caption{UI of our solution} \label{fig:ui_design}
%\end{figure}

% tez 1-2 paragraph ze to ewaluacja z rq.3 i  rq.4., why it is here, why important 
% rozpisac, że przeprowadzamy dwa case studies to validate the useability and applicapility of our system. pierwsze case study polega na xyz, drugie na pqr, przyklady wykorzystania a nie metodologia 
\section{Case Studies: Application of ARISE in Teacher Education}
In this section, we delve into the usability and applicability of the ARISE platform within the realm of teacher education to answer RQ.3 and 4. 
To this end, we design a scenario of a workshop, the SafeSchools Initiative, aimed at educating teachers about challenges and effective management of hate incidents, with semi-structured interviews conducted with a group of professional school teachers after such a workshop session with the system.
As part of this initiative, educators recognised a need for enhanced training in effectively managing and addressing these incidents when they occur. 
Teachers faced challenges in knowing how to respond appropriately when hate incidents occurred, partly due to a lack of specific training and the complex nature of these situations. Traditional training methods were time-consuming and limited in the number of training scenarios.
To this end, the ARISE platform was designed to simulate diverse contexts and provide personalised guidance using real-world examples.

Results of the study confirm the scarcity of precise guidelines for managing hate incidents in schools, although teachers feel confident in their ability to intervene effectively. Some teachers report that interventions should focus on educating students that words matter and the language they use can harm others even if employed humouristically. 
%This highlights the need for more detailed and structured guidelines and comprehensive training programs. 
Overall, the system's analysis of hate incidents is rated as satisfactory, although one individual noted overconfidence and overinterpretation in some cases as the system's weakness. 
Teachers acknowledged the representation of a multifaceted nature of hate, including the various underlying causes and impacts on students the system provides. At the same time, teachers raised the need for some kind of comparative analysis of hate instances against a ``legal norm'' of the country or school. We will explore adding such a module in future studies with the system as it necessitates the inclusion of a larger team of researchers who would evaluate the accuracy of this functionality. 

\section{Conclusions}
% jako future work konkluzje z evaluacji z user: ten legal module, nadal legal definition to nie jest ground truth, też "subiektywne" i system by pozwalał na porównanie tego co obecnie do danej definicji (ground-truth wymagało by wspólpracy z filozofami i prawnikami,  "prostsza" opcja dodać definicje HS z danego kraju / medium  )
% bez podawania liczby uzytkowników, tylko a group of ... to jako pilot study, dlatego nie podajemy stats 
% do appendixa zbierac reszte na artykul do czasopisma 
% cytowania naszych prac uwzględnić
% International Journal of Artificial Intelligence in Education

This study highlights the transformative potential of LLMs in enhancing teachers' skills of effective hate incident management. While the application of artificial intelligence in educational environments has been limited, our multi-agent LLM-based system demonstrates a promising approach to simulating realistic scenarios that educators may encounter. By addressing the complexities associated with hate speech—such as context dependence, varying definitions, and annotator disagreements—our system equips teachers with the necessary tools to navigate these challenges effectively. 
The pilot evaluation indicates that teachers gained a deeper understanding of the nuances of interpreting hate speech cases, increasing their skills in managing such sensitive situations. 
As educational institutions continue to strive for inclusivity and safety, integrating advanced AI technologies like the proposed ARISE platform can play a crucial role in preparing educators to confront and mitigate hate incidents proactively.

\section*{Acknowledgment}
We would like to acknowledge that the work reported in this paper has been supported in part by the Polish National Science Centre, Poland (Chist-Era IV) under grant 2022/04/Y/ST6/00001.

%\newpage
\bibliographystyle{splncs04}
\bibliography{biblio}

\begin{thebibliography}{10}
\providecommand{\url}[1]{\texttt{#1}}
\providecommand{\urlprefix}{URL }
\providecommand{\doi}[1]{https://doi.org/#1}

\bibitem{achiam2023gpt}
Achiam, J., Adler, S., Agarwal, S., Ahmad, L., Akkaya, I., Aleman, F.L., Almeida, D., Altenschmidt, J., Altman, S., Anadkat, S., et~al.: Gpt-4 technical report. arXiv preprint arXiv:2303.08774  (2023)

\bibitem{alshahrani2024review}
Alshahrani, K., Qureshi, R.J.: Review the prospects and obstacles of ai-enhanced learning environments: The role of chatgpt in education. International Journal of Modern Education and Computer Science  \textbf{16}(4),  71--86 (2024)

\bibitem{bilz2024teachers}
Bilz, L., Fischer, S.M., Kansok-Dusche, J., Wachs, S., Wettstein, A.: Teachers’ intervention strategies for handling hate-speech incidents in schools. Social Psychology of Education pp. 1--24 (2024)

\bibitem{cinkusz2024towardsLlmAugmentedMultiagentSystems}
Cinkusz, K., Chudziak, J.A.: Towards llm-augmented multiagent systems for agile software engineering. In: Proceedings of the 39th IEEE/ACM International Conference on Automated Software Engineering. p. 2476–2477. Association for Computing Machinery (2024), \url{https://doi.org/10.1145/3691620.3695336}

\bibitem{elsherief-etal-2021-latent}
ElSherief, M., Ziems, C., Muchlinski, D., Anupindi, V., Seybolt, J., De~Choudhury, M., Yang, D.: Latent hatred: A benchmark for understanding implicit hate speech. In: Proceedings of the 2021 Conference on Empirical Methods in Natural Language Processing. pp. 345--363. Association for Computational Linguistics, Online and Punta Cana, Dominican Republic (Nov 2021), \url{https://aclanthology.org/2021.emnlp-main.29}

\bibitem{harbarChudziakOxfordStyleDebatesLLMs}
Harbar, Y., Chudziak, J.A.: Simulating oxford-style debates with llm-based multi-agent systems. In: Proceedings of the ACIIDS 2025 : 17th Asian Conference on Intelligent Information and Database Systems (2025)

\bibitem{kansok2023systematic}
Kansok-Dusche, J., Ballaschk, C., Krause, N., Zei{\ss}ig, A., Seemann-Herz, L., Wachs, S., Bilz, L.: A systematic review on hate speech among children and adolescents: Definitions, prevalence, and overlap with related phenomena. Trauma, violence, \& abuse  \textbf{24}(4),  2598--2615 (2023)

\bibitem{kim2024leading}
Kim, J.: Leading teachers' perspective on teacher-ai collaboration in education. Education and Information Technologies  \textbf{29}(7),  8693--8724 (2024)

\bibitem{kostkaChudziakLogiccalReasoning2024}
Kostka, A., Chudziak, J.A.: Synergizing logical reasoning, knowledge management and collaboration in multi-agent llm system. In: Proceedings of the 38th Pacific Asia Conference on Language, Information and Computation. Association for Computational Linguistics (Dec 2024)

\bibitem{lewis2021retrievalaugmentedgeneration}
Lewis, P., Perez, E., Piktus, A., Petroni, F., Karpukhin, V., Goyal, N., Küttler, H., Lewis, M., tau Yih, W., Rocktäschel, T., Riedel, S., Kiela, D.: Retrieval-augmented generation for knowledge-intensive nlp tasks (2021), \url{https://arxiv.org/abs/2005.11401}

\bibitem{mathew2021HateXplain}
Mathew, B., Saha, P., Yimam, S.M., Biemann, C., Goyal, P., Mukherjee, A.: Hatexplain: A benchmark dataset for explainable hate speech detection. In: Proceedings of the AAAI conference on artificial intelligence. vol.~35, pp. 14867--14875 (2021)

\bibitem{mondal2023chatgpt}
Mondal, H., Marndi, G., Behera, J.K., Mondal, S.: Chatgpt for teachers: Practical examples for utilizing artificial intelligence for educational purposes. Indian Journal of Vascular and Endovascular Surgery  (2023)

\bibitem{oyekoya2021exploring}
Oyekoya, O., Urbanski, J., Shynkar, Y., Baksh, A., Etsaghara, M.: Exploring first-person perspectives in designing a role-playing vr simulation for bullying prevention: A focus group study. Frontiers in Virtual Reality  \textbf{2},  672003 (2021)

\bibitem{pitura2024vr}
Pitura, J., Kaplan-Rakowski, R., Asotska-Wierzba, Y.: The vr-ai--assisted simulation for content knowledge application in pre-service efl teacher training. TechTrends pp. 1--11 (2024)

\bibitem{raiaanReviewOfLLMs2024}
Raiaan, M.A.K., Mukta, M.S.H., Fatema, K., Fahad, N.M., Sakib, S., Mim, M.M.J., Ahmad, J., Ali, M.E., Azam, S.: A review on large language models: Architectures, applications, taxonomies, open issues and challenges. IEEE Access  \textbf{12},  26839--26874 (2024). \doi{10.1109/ACCESS.2024.3365742}

\bibitem{ross2016measuring}
Ross, B., Rist, M., Carbonell, G., Cabrera, B., Kurowsky, N., Wojatzki, M.: Measuring the reliability of hate speech annotations: The case of the european refugee crisis. In: 3rd Workshop on Natural Language Processing for Computer-Mediated Communication/Social Media. pp.~6--9. Ruhr-Universitat Bochum (2016)

\bibitem{sap2021annotators}
Sap, M., Swayamdipta, S., Vianna, L., Zhou, X., Choi, Y., Smith, N.A.: Annotators with attitudes: How annotator beliefs and identities bias toxic language detection. In: Carpuat, M., de~Marneffe, M.C., Meza~Ruiz, I.V. (eds.) Proceedings of the 2022 Conference of the North American Chapter of the Association for Computational Linguistics: Human Language Technologies. pp. 5884--5906. Association for Computational Linguistics, Seattle, United States (Jul 2022). \doi{10.18653/v1/2022.naacl-main.431}, \url{https://aclanthology.org/2022.naacl-main.431/}

\bibitem{zhoSurveyOfLLMsAgents2023}
Zhao, P., Jin, Z., Cheng, N.: An in-depth survey of large language model-based artificial intelligence agents. arXiv:2309.14365v1 [cs.CL] (9 2023)

\bibitem{stefuanitua2021hate}
Ștef{\u{a}}niț{\u{a}}, O., Buf, D.M.: Hate speech in social media and its effects on the {LGBT} community: A review of the current research. Romanian Journal of Communication and Public Relations  \textbf{23}(1),  47--55 (2021)

\end{thebibliography}

\end{document}